\documentclass{jpsj3}
\usepackage{txfonts}
\usepackage{color}

\title{High-Frequency Analysis of Effective Interactions and Bandwidth for Transient States after Monocycle Pulse Excitation of Extended Hubbard Model}

\author{Kenji Yonemitsu\thanks{E-mail: kxy@phys.chuo-u.ac.jp}}
\inst{Department of Physics, Chuo University, Bunkyo, Tokyo 112-8551, Japan} 

\abst{ Using a high-frequency expansion in periodically driven extended Hubbard models, where the strengths and ranges of density-density interactions are arbitrary, we obtain the effective interactions and bandwidth, which depend sensitively on the polarization of the driving field. Then, we numerically calculate modulations of correlation functions in a quarter-filled extended Hubbard model with nearest-neighbor interactions on a triangular lattice with trimers after monocycle pulse excitation. We discuss how the resultant modulations are compatible with the effective interactions and bandwidth derived above on the basis of their dependence on the polarization of photoexcitation, which is easily accessible by experiments. Some correlation functions after monocycle pulse excitation are consistent with the effective interactions, which are weaker or stronger than the original ones. However, the photoinduced enhancement of anisotropic charge correlations previously discussed for the three-quarter-filled organic conductor $\alpha$-(bis[ethylenedithio]-tetrathiafulvalene)$_2$I$_3$ [$\alpha$-(BEDT-TTF)$_2$I$_3$] in the metallic phase is not fully explained by the effective interactions or bandwidth, which are derived independently of the filling. }

\begin{document}
\maketitle

\section{Introduction}
Nonequilibrium properties of quantum many-body systems have received much attention, which can lead to advances in their real-time and coherent manipulation. Motivated by experiments on ultracold atomic gases, interaction quench has been discussed theoretically in different contexts.\cite{eckstein_prl08,moeckel_prl08,eckstein_prl09,moeckel_anphys09,poletti_pra11,mentink_prl14,nessi_prl14} For many-electron systems in solids, periodic driving is achievable, including photoexcitation. Electromagnetic fields are often incorporated into the Peierls phase multiplied by transfer integrals. In particular, for continuous waves, long-time dynamics compared with the period of the oscillating field has been discussed to develop the concept of dynamical localization.\cite{dunlap_prb86,grossmann_prl91,kayanuma_pra94} The corresponding effective Hamiltonian is simply the time average of the time-dependent Hamiltonian and is regarded as the lowest-order ($ \propto \omega^{0} $) term in a high-frequency ($ \omega $) expansion for an effective Hamiltonian in the framework of quantum Floquet theory.\cite{rahav_pra03,mananga_jcp11,goldman_prx14,eckardt_njp15,itin_prl15,bukov_ap15,mikami_prb16} For instance, a Floquet topological insulator can be discussed in the second-lowest order ($ \propto \omega^{-1} $).\cite{oka_prb09} In the next order ($ \propto \omega^{-2} $), local interactions are modulated.\cite{rahav_pra03,mananga_jcp11,goldman_prx14,eckardt_njp15,itin_prl15,bukov_ap15,mikami_prb16} Bearing many-electron systems in solids in mind, we first consider extended Hubbard models, where the strengths and ranges of density-density interactions are arbitrary. 

Most photoinduced phase transitions are triggered by a pulse of light.\cite{koshigono_jpsj06,yonemitsu_pr08,basov_rmp11,nicoletti_aop16} As the pulse width decreases, the time resolution is improved and the instantaneous field amplitude increases. At the same time, oscillating electric fields are viewed as coherently driving many electrons.\cite{kawakami_prl10,matsubara_prb14} Recently, the optical freezing  of charge motion\cite{ishikawa_ncomms14} and the photoinduced suppression of conductivity\cite{fukaya_ncomms15} have been observed. For the former, the similarity to dynamical localization was pointed out, although dynamical localization is a continuous-wave-induced phenomenon. 

Various similarities between continuous-wave- and pulse-induced phenomena are known. A negative-temperature state is produced in both cases if the electric field amplitude is large and satisfies a certain condition.\cite{tsuji_prl11,tsuji_prb12,yonemitsu_jpsj15,yanagiya_jpsj15} A sudden application of a continuously oscillating weak electric field to the half-filled Hubbard model immediately decreases the double occupancy.\cite{tsuji_prl11} Similar behavior has also been reported in a one-dimensional Bose-Hubbard model.\cite{poletti_pra11} Because this early-stage dynamics does not depend on whether the field continues to be applied or not, in both the continuous-wave and pulse cases the transient state behaves as if the interaction strength were increased relative to the bandwidth. The transition from a charge-ordered insulator phase to a Mott insulator phase in the quasi-two-dimensional metal complex Et$_2$Me$_2$Sb[Pd(dmit)$_2$]$_2$ (dmit = 1,3-dithiol-2-thione-4,5-dithiolate)\cite{ishikawa_prb09} can theoretically be controlled  by suppressing the effective transfer integrals in both cases.\cite{nishioka_jpsj14} The similarity between continuous-wave- and pulse-induced phenomena has also been discussed for a one-dimensional transverse Ising model.\cite{ono_prb16} 

In practice, the application of laser pulses is more advantageous than that of continuous waves for ultrafast collective phenomena that become possible only when the electric field amplitude is large. The optical freezing of charge motion\cite{ishikawa_ncomms14} is indeed one such phenomenon. Theoretically, for pulse-induced phenomena in quantum many-body systems, only numerical approaches have so far been employed. In this context, an analytic approach will be useful if similarities are empirically found between continuous-wave- and pulse-induced phenomena, even if it is basically developed for continuous waves. In this study, we employ a high-frequency expansion to obtain an effective Hamiltonian in the framework of quantum Floquet theory and discuss the behavior generally expected after periodic driving. Then, we tentatively use it to analyze pulse-induced transient states. If transient states including those similar to dynamically localized states survive for a while, they may be described by the effective Hamiltonian, which has renormalized transfer integrals and interactions. 

We will compare states expected by the effective Hamiltonian and monocycle-pulse-induced transient states, which are numerically obtained by solving a time-dependent Schr\"odinger equation. Although the photoinduced enhancement of anisotropic charge correlations previously discussed for the 3/4-filled organic conductor $\alpha$-(bis[ethylenedithio]-tetrathiafulvalene)$_2$I$_3$ [$\alpha$-(BEDT-TTF)$_2$I$_3$] in the metallic phase\cite{yonemitsu_jpsj17a} is not reproduced by the effective Hamiltonian, it is shown to be generally useful when we roughly expect transient states after monocycle pulse excitation. 

\section{High-Frequency Approximation for Periodically Driven, Extended Hubbard Models \label{sec:hfe}}
In this section, we do not specify the dimension or lattice structure (i.e., network of transfer integrals) and generally consider extended Hubbard models, where the strengths and ranges of density-density interactions are arbitrary, 
\begin{equation}
H = 
\sum_{i,j(\neq i),\sigma}
t_{ij}  c^\dagger_{i,\sigma} c_{j,\sigma} 
+\frac{1}{2}\sum_{i,\sigma} U_{i} n_{i,\sigma} n_{i,-\sigma}
+\frac{1}{2}\sum_{i,j(\neq i),\sigma,\tau} V_{ij} n_{i,\sigma} n_{j,\tau}
\;, \label{eq:model}
\end{equation}
where $ c^\dagger_{i\sigma} $ creates an electron with spin $ \sigma $ at site $ i $ and $ n_{i\sigma} $=$ c^\dagger_{i\sigma} c_{i\sigma} $. The parameters $ t_{ij} $ and $ V_{ij} $ denote the transfer integral and the intersite repulsion, respectively, between sites $ i $ and $ j $, and $ U_{i} $ denotes the on-site repulsion at site $ i $. Photoexcitation is introduced through the Peierls phase, 
\begin{equation}
c_{i,\sigma}^\dagger c_{j,\sigma} \rightarrow
\exp \left[
\frac{ie}{\hbar c} \mbox{\boldmath $r$}_{ij} \cdot \mbox{\boldmath $A$}(t)
\right] c_{i,\sigma}^\dagger c_{j,\sigma}
\;, \label{eq:photo_excitation}
\end{equation}
with $ \mbox{\boldmath $r$}_{ij}=\mbox{\boldmath $r$}_j-\mbox{\boldmath $r$}_i $. In this section, we consider the time-dependent vector potential for a continuous wave, 
\begin{equation}
\mbox{\boldmath $A$} (t) = -\frac{c\mbox{\boldmath $F$}}{\omega} \sin (\omega t)
\;, \label{eq:continuous_wave}
\end{equation}
where $ \omega $ is the frequency and $ \mbox{\boldmath $F$} $ describes the amplitude and polarization of the electric field. When we substitute Eq.~(\ref{eq:continuous_wave}) into Eq.~(\ref{eq:photo_excitation}), we obtain the Peierls phase factor, which is expanded as 
\begin{equation}
\exp \left[ i\frac{ea_{ij}F}{\hbar\omega} 
\cos(\phi_{ij}-\theta) \sin \omega t \right] = \sum_{m=-\infty}^{\infty} 
J_m(ij) \exp( i m \omega t )
\;, \label{eq:bessel_expansion}
\end{equation}
where $ J_m(ij) $ denotes
\begin{equation}
J_m(ij) \equiv J_m \left( \frac{ea_{ij}F}{\hbar\omega} 
\cos(\phi_{ij}-\theta) \right)
\;, \label{eq:jm_ij}
\end{equation}
with $ J_m(x) $ being the $m$th-order Bessel function,  $ a_{ij} = \mid \mbox{\boldmath $r$}_i-\mbox{\boldmath $r$}_j \mid $, and $ \phi_{ij} $ is the angle between $ \mbox{\boldmath $r$}_i-\mbox{\boldmath $r$}_j $ (not $ \mbox{\boldmath $r$}_j-\mbox{\boldmath $r$}_i $) and a reference axis. Note that $ \phi_{ji} = \phi_{ij} + \pi \pmod{2\pi} $; thus, $ \cos(\phi_{ji}-\theta) = - \cos(\phi_{ij}-\theta) $. Note also that $ J_m(-x)= J_{-m}(x)= (-1)^m J_m(x) $. 

For continuous waves, as long as the time evolution is considered in a stroboscopic manner in steps of the period $ T = 2\pi/\omega $, the stroboscopic time evolution is described by a time-independent effective Hamiltonian. The effective Hamiltonian is approximately derived by a high-frequency  expansion.\cite{rahav_pra03,mananga_jcp11,goldman_prx14,eckardt_njp15,itin_prl15,bukov_ap15,mikami_prb16} We follow Ref.~\citen{eckardt_njp15}, which employs degenerate perturbation theory in the extended Floquet Hilbert space, and use its notations. In the lowest order, we have 
\begin{equation}
H_{F}^{(1)} = H_0 + H_{\mbox{int}}
\;, \label{eq:expansion_1}
\end{equation}
where 
\begin{equation}
H_0  = \sum_{i,j(\neq i),\sigma} 
t_{ij} J_0(ij) c^\dagger_{i,\sigma} c_{j,\sigma} 
\; \label{eq:0th_bessel}
\end{equation}
and 
\begin{equation}
H_{\mbox{int}} = \frac{1}{2}\sum_{i,\sigma} U_{i} n_{i,\sigma} n_{i,-\sigma}
+\frac{1}{2}\sum_{i,j(\neq i),\sigma,\tau} V_{ij} n_{i,\sigma} n_{j,\tau}
\;. \label{eq:interaction}
\end{equation}
In the second-lowest order, we have 
\begin{equation}
H_{F}^{(2)} = \sum_{m \neq 0} \frac{H_{m}H_{-m}}{m \hbar \omega}
\;, \label{eq:expansion_2}
\end{equation}
where 
\begin{equation}
H_m  = \sum_{i,j(\neq i),\sigma} 
t_{ij} J_m(ij) c^\dagger_{i,\sigma} c_{j,\sigma} 
\;. \label{eq:mth_bessel}
\end{equation}
In the next order, we obtain 
\begin{equation}
H_{F}^{(3)} = \sum_{m \neq 0} \left( 
\frac{[ H_{-m},[ H_{0}+H_{\mbox{int}}, H_{m}]]}{2(m \hbar \omega)^2}
+ \sum_{m' \neq 0,m} 
\frac{[ H_{-m'},[ H_{m'-m}, H_{m}]]}{3mm'(\hbar \omega)^2} \right)
\;; \label{eq:expansion_3}
\end{equation}
thus, we define 
\begin{equation}
H_{F,\mbox{int}}^{(3)} = \sum_{m \neq 0} 
\frac{[ H_{-m},[ H_{\mbox{int}}, H_{m}]]}{2(m \hbar \omega)^2}
\;, \label{eq:interaction_3}
\end{equation}
as in Ref.~\citen{eckardt_njp15}. 

We decompose the interaction term 
\begin{equation}
H_{\mbox{int}} = H_{U} + H_{V}
\; \label{eq:interaction_UV}
\end{equation}
into 
\begin{equation}
H_{U} = \frac{1}{2}\sum_{i,\sigma} U_{i} n_{i,\sigma} n_{i,-\sigma}
\; \label{eq:interaction_U}
\end{equation}
and 
\begin{equation}
H_{V} = \frac{1}{2}\sum_{i,j(\neq i),\sigma,\tau} V_{ij} n_{i,\sigma} n_{j,\tau}
\; \label{eq:interaction_V}
\end{equation}
to obtain
\begin{equation}
H_{F,\mbox{int}}^{(3)} = \sum_{m \neq 0} 
\frac{[ H_{-m},[ H_{U}, H_{m}]] + [ H_{-m},[ H_{V}, H_{m}]]}{2(m \hbar \omega)^2}
\;. \label{eq:interaction_3UV}
\end{equation}
The double commutators in Eq.~(\ref{eq:interaction_3UV}) are calculated in the Appendix. Substituting Eqs.~(\ref{eq:density_density_1}) and (\ref{eq:density_density_2}) into Eq.~(\ref{eq:H_mVm}) and similar terms, which are obtained by setting $l_0=l_2$ or $l_0=l_3$ in Eq.~(\ref{eq:density_density_1}), into Eq.~(\ref{eq:H_mUm}), and further substituting the resultant Eqs.~(\ref{eq:H_mUm}) and (\ref{eq:H_mVm}) into Eq.~(\ref{eq:interaction_3UV}), we obtain the modulations of on-site and intersite repulsive interactions, which are 
\begin{equation}
\delta U_{i} = \sum_{m=1}^{\infty} \frac{4}{(m \hbar \omega)^2} 
\sum_{j} ( V_{ij}-U_{i} ) t_{ij}^2 J_m^2(ij) 
\;, \label{eq:delta_U}
\end{equation}
and 
\begin{eqnarray}
\delta V_{ij} & & = \sum_{m=1}^{\infty} \frac{4}{(m \hbar \omega)^2} 
\left[ \left( \frac{U_{i}+U_{j}}{2} -V_{ij} \right) t_{ij}^2 J_m^2(ij) \right.
\nonumber \\ & & \left.
+\sum_{k} (V_{kj}-V_{ij}) t_{ik}^2 J_m^2(ik) \right]
\;. \label{eq:delta_V}
\end{eqnarray}
Since the modulations $ \delta U_{i} $ and $ \delta V_{ij} $ originate from the double commutators in Eq.~(\ref{eq:interaction_3UV}), they depend sensitively on $ \theta $. 

So far, we have distinguished on-site ($ U_{i} $) and intersite ($ V_{ij} $) repulsions. If we set $ V_{ii} = U_{i} $ and sum over sites $ i $ and $ j $ irrespective of whether they are different or not in the third term and ignore the second term in Eq.~(\ref{eq:model}), the difference between the resultant Hamiltonian and Eq.~(\ref{eq:model}) is a one-body term that becomes a constant when $ U_{i} $ is independent of $ i $ because the total number of electrons $ N_e $ is conserved. This fact is useful in checking formulae. In fact, Eqs.~(\ref{eq:delta_U}) and (\ref{eq:delta_V}) are consistent: Eq.~(\ref{eq:delta_U}) is a special case of Eq.~(\ref{eq:delta_V}) with $ i = j $. If we consider the case where $ U_{i}=\gamma $ and $ V_{ij}=\gamma $ for all $i$ and $j$ (thus, $H_{\mbox{int}} = \frac{1}{2} \gamma N_e^2 $), the interaction is conserved and independent of time, so that $ U_{i} $ and $ V_{ij} $ are not modified at all. Equations (\ref{eq:delta_U}) and (\ref{eq:delta_V}) satisfy this condition, as easily checked. Note that Eqs.~(\ref{eq:delta_U}) and (\ref{eq:delta_V}) are independent of the filling or the system size. Therefore, the effective interactions will not be able to describe any phenomena that are sensitive to the filling or the system size. 

As is evident from the Appendix, most of the terms in $ H_{F,\mbox{int}}^{(3)} $ are not interactions between site-diagonal densities, but they bring about electron transfers. As a consequence, continuous-wave-induced changes in the electronic state will generally tend to homogenize the electron distribution if the charge is initially disproportionated. When we focus on the interaction terms between site-diagonal densities, the modulations of the interaction strengths $ \delta U_{i} $ [Eq.~(\ref{eq:delta_U})] and $ \delta V_{ij} $ [Eq.~(\ref{eq:delta_V})] have contributions from themselves with negative signs (thus weakening the effective $ U_{i} $ and $ V_{ij} $ by themselves) and contributions from other strengths with positive signs (thus are enhanced by all other strengths with the same signs). This implies that large interaction parameters become small while small interaction parameters become large, thus averaging themselves out. Indeed, in the special case where all the interaction parameters are equal ($ U_{i}=\gamma $ and $ V_{ij}=\gamma $ for all $i$ and $j$), the modulation is absent (all the interaction parameters are already averaged out). It is natural to assume that the modulations approach the special form of the interaction $ \frac{1}{2} \gamma N_e^2 $, which is actually equivalent to no interaction because $ N_e $ is a constant. 

In realistic cases, interaction strengths are zero or very small between distant sites, large between neighboring sites, and largest on a single site. Therefore, the effective Hamiltonian will possess weaker interactions between neighbors and stronger interactions between distant sites than the original one. If there is anisotropy in intersite interactions between neighboring sites, the anisotropy will be suppressed and the effective Hamiltonian will acquire isotropic interactions. Note that the rates of the modulations are governed by the square of a transfer integral multiplied by a corresponding non-zeroth-order Bessel function divided by $ \omega $, as shown in Eqs.~(\ref{eq:delta_U}) and (\ref{eq:delta_V}), leading to their sensitivity to $ \theta $. 

\section{Correlations after Monocycle Pulse Excitation}

\subsection{Extended Hubbard model on triangular lattice}
\begin{figure}
\includegraphics[height=14.3cm]{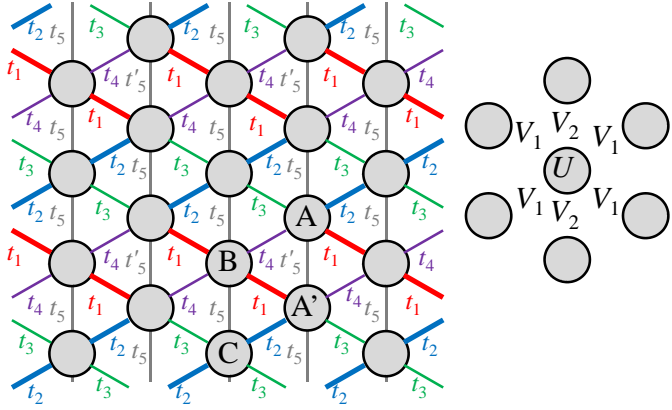}
\caption{(Color online) 
Triangular lattice with linear trimers linked by $t_1$. 
\label{fig:latt_str}}
\end{figure}

In this section, we specify the network of transfer integrals and the strengths and ranges of density-density interactions. We use a quarter-filled extended Hubbard model with on-site and nearest-neighbor repulsions on the triangular lattice with linear trimers shown in Fig.~\ref{fig:latt_str}, which was previously employed to study the mechanism for the photoinduced tendency toward charge localization.\cite{yonemitsu_jpsj17a} 
The triangular lattice we consider here consists of equilateral triangles, where the distance between neighboring sites is denoted by $ a $, and has inversion symmetry. The use of this model facilitates a comparison of the high-frequency expansion of the effective Hamiltonian and numerical results. 

For the transfer integrals $ t_{ij} $, we use $t_1=-0.14$, $t_2=-0.13$, $t_3=-0.02$, $t_4=-0.06$, $t_5=0.03$, and $t_5'=-0.03$ in Fig.~\ref{fig:latt_str}, as before.\cite{yonemitsu_jpsj17a} For the on-site repulsion, we consider $ U_{i} = U $ for all $ i $ and use $U=0.8$ unless stated otherwise. For the intersite repulsions, we take only nearest-neighbor Coulomb repulsions and set $V_{ij}=V_1$ for $ \mbox{\boldmath $r$}_{ij}$ not being parallel to the vertical axis and $V_{ij}=V_2$ for $ \mbox{\boldmath $r$}_{ij}$ being parallel to the vertical axis, as shown in Fig.~\ref{fig:latt_str}, and use $V_1=0.3$ unless stated otherwise. 

\begin{figure}
\includegraphics[height=12.0cm]{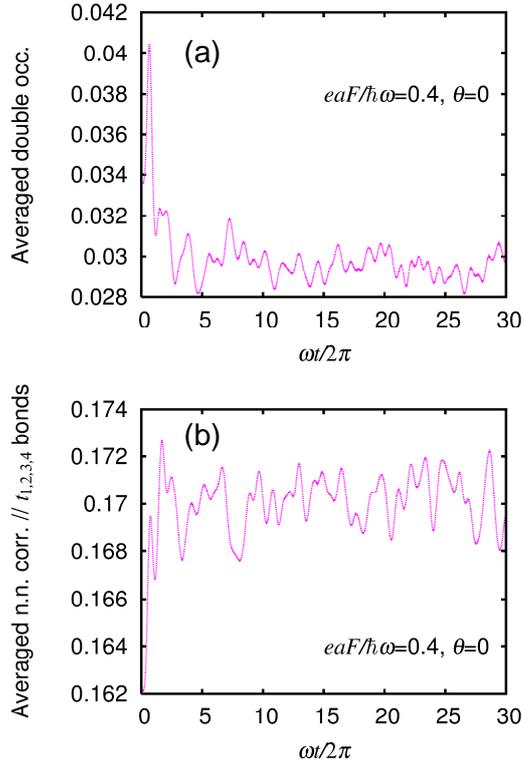}
\caption{(Color online) 
Time evolution of (a) spatially averaged double occupancy $\langle n_{i\uparrow} n_{i\downarrow} \rangle$ of Fig.~\ref{fig:avdo}(a) shown later and (b) spatially averaged nearest-neighbor density-density correlation $\langle n_i n_j \rangle$ for non-vertical $ \mbox{\boldmath $r$}_{ij}$ of Fig.~\ref{fig:avnn1234}(a) shown later, both with $ eaF/(\hbar\omega) $=0.4 and $\theta$=0. 
\label{fig:time_evol}}
\end{figure}

The initial state is the ground state obtained by the exact diagonalization method for the 16-site system with periodic boundary conditions. The time-dependent vector potential in the Peierls phase is now set to be\cite{yonemitsu_jpsj15,yanagiya_jpsj15} 
\begin{equation}
\mbox{\boldmath $A$} (t) = \frac{c\mbox{\boldmath $F$}}{\omega} \left[ \cos (\omega t)-1 \right] 
\theta (t) \theta \left( \frac{2\pi}{\omega}-t \right)
\;, \label{eq:monocycle_pulse}
\end{equation}
with $ \mbox{\boldmath $F$}=F(\cos \theta,\sin \theta) $, where $F$ is the amplitude of the electric field and $\theta$ is the angle between the field and the horizontal axis. As in the previous paper,\cite{yonemitsu_jpsj17a} we use $ \omega=0.8 $. The time-dependent Schr\"odinger equation is numerically solved by expanding the exponential evolution operator with a time slice $ dt $=0.02 to the 15th order and by checking the conservation of the norm.\cite{yonemitsu_prb09} The time average $\langle \langle Q \rangle \rangle $ of a quantity $Q$ is calculated by 
\begin{equation}
\langle \langle Q \rangle \rangle =
\frac{1}{t_w} \int_{t_s}^{t_s+t_w} 
\langle \Psi (t) \mid Q \mid \Psi (t) \rangle dt
\;, \label{eq:time_average}
\end{equation}
with $ t_s=5 T $ and $ t_w=5 T $, where $ T $ is the period $ T=2\pi/\omega $. 
Figures \ref{fig:time_evol}(a) and \ref{fig:time_evol}(b) represent the time evolutions of quantities shown later in Figs.~\ref{fig:avdo}(a) and \ref{fig:avnn1234}(a), respectively. 
For $ eaF/(\hbar\omega) $=0.4 with $ \hbar\omega $=0.8 eV and $ a $=5.9 {\AA} (4.6 {\AA}), which is close to the intermolecular distance along ``$b$'' (``$a$'') bonds of $\alpha$-(BEDT-TTF)$_2$I$_3$, the field amplitude corresponds to about $F$=5.4 MV/cm (7.0 MV/cm) that is available in recent experiments.\cite{ishikawa_ncomms14} Since $ \omega $=0.8 corresponds to the largest model parameter ($U$=0.8), oscillations appearing after photoexcitation [$ \omega t/(2 \pi) = t/T > 1 $] have frequencies that are comparable to $ \omega $. In spite of the presence of such oscillations, the transient quantities are significantly different from the corresponding initial values. Thus, only time averages are shown below for the transient states. 

Although details are discussed later, here we briefly comment on the fact that the behavior shown in Fig.~\ref{fig:time_evol}(a) is similar to dynamical localization and its time average appears consistent with the lowest-order effective Hamiltonian. Similar numerical results have been reported for Fermi-Hubbard\cite{tsuji_prl11} and Bose-Hubbard\cite{poletti_pra11} models after a sudden application of {\it continuously oscillating} fields. For a one-dimensional Bose-Hubbard model, field-induced states have been analyzed for different $ \omega $ values, field amplitudes, and time scales.\cite{poletti_pra11} For high frequencies ($ \omega \gg U $), dynamical localization is observed. For a resonant excitation ($ \omega = U $), the behavior is similar to dynamical localization only for the first few cycles and quickly deviates from it owing to the resonant absorption of energy. This continuous-wave-induced behavior is consistent with a recently proved theorem,\cite{mori_prl16,kuwahara_anphys16,abanin_prb17,abanin_arx} which shows that the time evolution under a periodically driven Hamiltonian is close to that under a truncated Floquet Hamiltonian for a $ \omega $-dependent time scale. For high frequencies, the time scale is exponentially long. As $ \omega $ decreases, the time scale is shortened. Eventually, systems are generally expected to reach a steady state of infinite temperature, although several exceptions are known. On the other hand, this paper deals with short-time behaviors after {\it monocycle pulse} excitation, so that their $ \omega $-dependence is much weaker than that after the application of {\it continuously oscillating} fields. 

\subsection{Expectation from effective Hamiltonian}
\begin{figure}
\includegraphics[height=12.0cm]{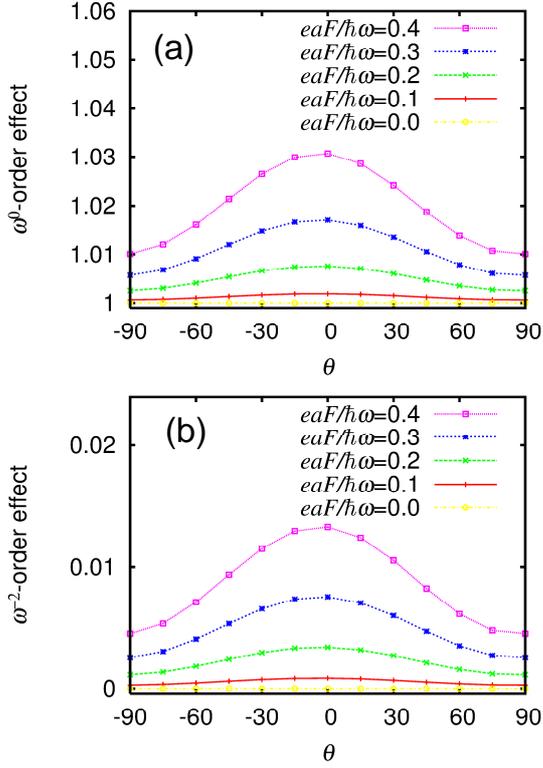}
\caption{(Color online) 
(a) Inverse of bandwidth effectively reduced by continuous-wave excitation relative to original value, Eq.~(\ref{eq:0th_order}), and (b) magnitude of continuous-wave-induced modulation of interactions relative to original values, Eq.~(\ref{eq:2nd_order}), as functions of polarization of photoexcitation $\theta$. 
\label{fig:flq_hfe}}
\end{figure}

Before showing numerical results for time averages, we discuss what is expected from the effective Hamiltonian obtained in Sect.~\ref{sec:hfe}. The bandwidth is a scale of the kinetic energy. Without interactions and when $ \mid t_1 \mid $ and $ \mid t_2 \mid $ are much larger than the other transfer integrals, the bandwidth $ W $ is proportional to $ \sqrt{t_1^2+t_2^2} $. In the lowest order ($ \propto \omega^{0} $) of the high-frequency expansion, $ t_1 $ is renormalized to be $ t_1 J_0 (ij\in t_1 \mbox{bond}) $, where ``$ ij\in t_1 \mbox{bond} $'' means that the argument of the Bessel function is that of Eq.~(\ref{eq:jm_ij}) with sites $i$ and $j$ being linked by $ t_1 $. In the same manner, $ t_2 $ is renormalized to be $ t_2 J_0 (ij\in t_2 \mbox{bond}) $. The renormalized bandwidth $ W + \delta W $ is proportional to $ \sqrt{t_1^2 J_0^2(ij\in t_1 \mbox{bond}) + t_2^2 J_0^2(ij\in t_2 \mbox{bond})} $. Then, in the lowest order, the ratio of any interaction to the bandwidth is increased by a factor of 
\begin{equation}
\frac{W}{W + \delta W} \simeq 
\frac{\sqrt{t_1^2 + t_2^2}}
{\sqrt{t_1^2 J_0^2(ij\in t_1 \mbox{bond}) + t_2^2 J_0^2(ij\in t_2 \mbox{bond})}}
\;, \label{eq:0th_order}
\end{equation}
whose dependence on the polarization of photoexcitation $ \theta $ is shown in Fig.~\ref{fig:flq_hfe}(a) 

Interaction parameters are renormalized by non-zeroth-order Bessel functions in the order of $ \omega^{-2} $. For small electric field amplitudes, the arguments of Bessel functions are small, so that the renormalization is dominated by first-order Bessel functions. Quantitatively, how $ U_{i} $ and $ V_{ij} $ are renormalized depends on the site indices in a rather complicated manner. When $ \mid t_1 \mid $ and $ \mid t_2 \mid $ are much larger than the other transfer integrals, they give most of the contributions. A rough estimation leads to 
\begin{equation}
\left| \frac{\delta V}{V} \right| \sim \frac{4}{(\hbar \omega)^2} \left[
2t_1^2 J_1^2(ij\in t_1 \mbox{bond}) + 2t_2^2 J_1^2(ij\in t_2 \mbox{bond}) \right]
\;, \label{eq:2nd_order}
\end{equation}
since sites $i$ and $j$ are linked by two large transfer integrals. Depending on the neighboring $ U_{i} $, $ U_{j} $, and $ V_{kj} $ in Eqs.~(\ref{eq:delta_U}) and (\ref{eq:delta_V}), $ \delta U_{i} $ and $ \delta V_{ij} $ can be positive or negative and they have different numerical factors, so that we here show only the relative magnitude and ignore the numerical factor. Thus, the estimation above is very rough. However, these details do not depend on $ \theta $, so the dependence of the right-hand side of Eq.~(\ref{eq:2nd_order}) on $ \theta $, which is shown in Fig.~\ref{fig:flq_hfe}(b), will be useful for various comparisons. 

The quantities in both Eqs.~(\ref{eq:0th_order}) and (\ref{eq:2nd_order}) reach a maximum around $ \theta = 0 $ (i.e., for polarization nearly parallel to the horizontal axis). When the argument is small, zeroth-order Bessel functions quadratically decrease from unity, so that any quantity with zeroth-order Bessel function(s) in its denominator quadratically increases, and the square of a first-order Bessel function quadratically increases from zero. Thus, the behaviors of the quantities in Eqs.~(\ref{eq:0th_order}) and (\ref{eq:2nd_order}) are similar. The smallness of the quantity shown in Fig.~\ref{fig:flq_hfe}(b) suggests that the high-frequency expansion converges rapidly for the field amplitudes used here. 

\subsection{Numerical results for time-averaged correlations}
\begin{figure}
\includegraphics[height=12.0cm]{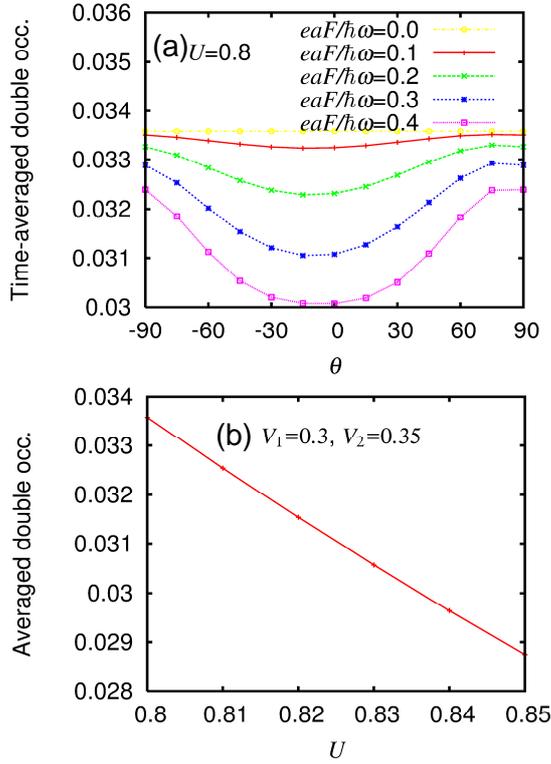}
\caption{(Color online) 
(a) Spatially and temporally averaged double occupancy 
$\langle \langle n_{i\uparrow} n_{i\downarrow} \rangle \rangle$ 
as a function of polarization of photoexcitation $\theta$ 
for $U$=0.8, $V_1$=0.3, $V_2$ = 0.35, and different field amplitudes $ eaF/(\hbar\omega) $. 
(b) Spatially averaged double occupancy of ground state 
$\langle n_{i\uparrow} n_{i\downarrow} \rangle$ 
as a function of $U$ for $V_1$=0.3 and $V_2$=0.35. 
\label{fig:avdo}}
\end{figure}

Now we show numerical results for time averages of correlation functions after monocycle pulse excitation. For $V_2=0.35$, whose case was investigated in detail in the previous paper,\cite{yonemitsu_jpsj17a} the spatially and temporally averaged double occupancy $\langle \langle n_{i\uparrow} n_{i\downarrow} \rangle \rangle$ is shown in Fig.~\ref{fig:avdo}(a) as a function of the polarization of photoexcitation $ \theta $. 
It decreases as if the on-site repulsion $U$ were transiently increased relative to the bandwidth after photoexcitation, as previously reported. Its $ \theta $ dependence is similar to that in Fig.~\ref{fig:flq_hfe}(a). To investigate whether the decrease in  $\langle \langle n_{i\uparrow} n_{i\downarrow} \rangle \rangle$ can be explained quantitatively by the increased ratio of the on-site repulsion $ U $ to the renormalized bandwidth, we vary $ U $ and calculate the spatially averaged double occupancy of the ground state $\langle n_{i\uparrow} n_{i\downarrow} \rangle$, as shown in Fig.~\ref{fig:avdo}(b). For $ \theta = 0 $ and $ eaF/(\hbar\omega) = 0.4 $ in Fig.~\ref{fig:avdo}(a), $\langle \langle n_{i\uparrow} n_{i\downarrow} \rangle \rangle$ is about 0.030. To reproduce this value in the ground state, we need to increase $ U $ by 4\%, as shown in Fig.~\ref{fig:avdo}(b). Figure~\ref{fig:flq_hfe}(a) shows that the ratio of the on-site repulsion $ U $ to the renormalized bandwidth is increased by about 3\% for $ \theta = 0 $ and $ eaF/(\hbar\omega) = 0.4 $. These values are comparable. 

\begin{figure}
\includegraphics[height=12.0cm]{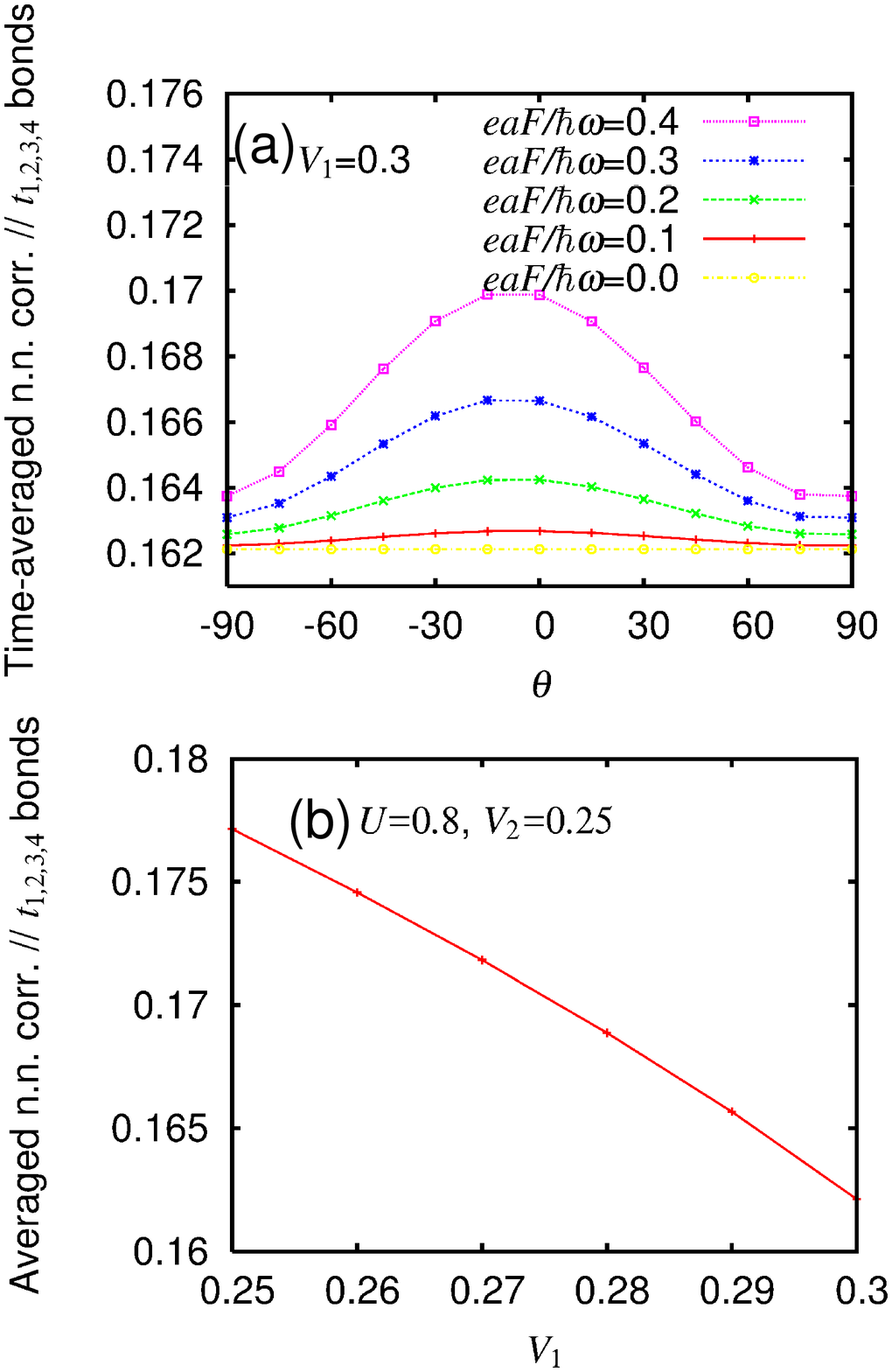}
\caption{(Color online) 
(a) Spatially and temporally averaged nearest-neighbor density-density correlation $\langle \langle n_i n_j \rangle \rangle$ for non-vertical $ \mbox{\boldmath $r$}_{ij}$ as a function of polarization of photoexcitation $\theta$ for $U$=0.8, $V_1$=0.3, $V_2$ = 0.25, and different field amplitudes $ eaF/(\hbar\omega) $. 
(b) Spatially averaged nearest-neighbor density-density correlation of ground state $\langle n_i n_j \rangle$ for non-vertical $ \mbox{\boldmath $r$}_{ij}$ as a function of $V_1$ for $U$=0.8 and $V_2$=0.25. 
\label{fig:avnn1234}}
\end{figure}

\begin{figure}
\includegraphics[height=24.0cm]{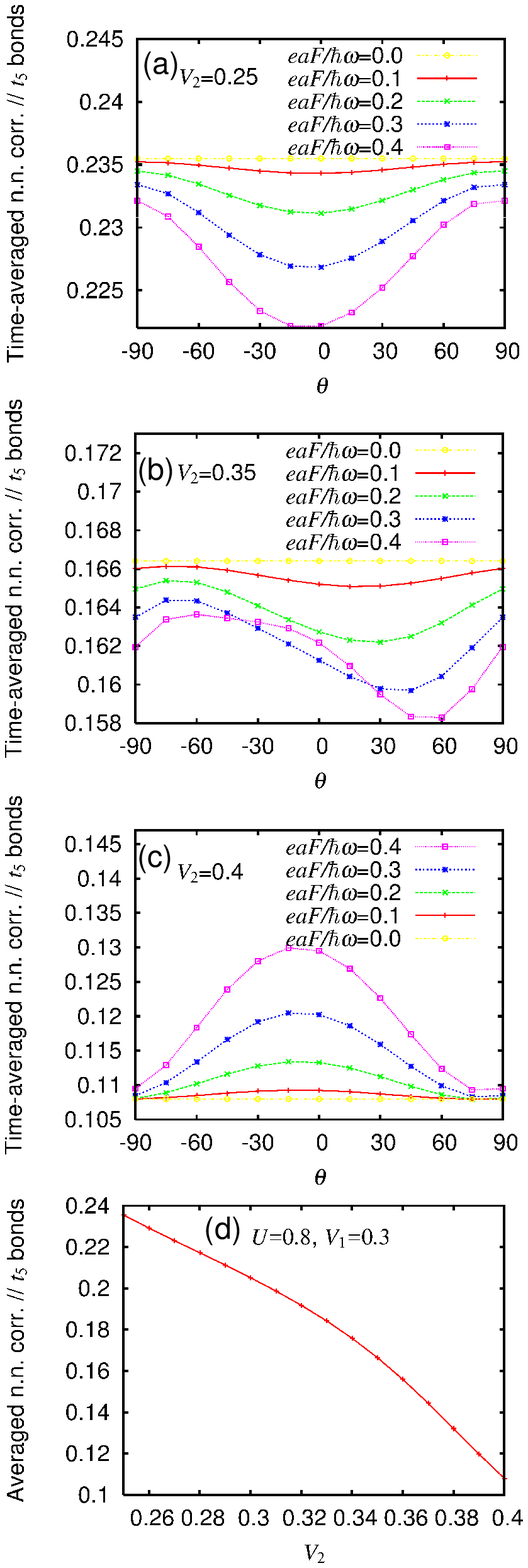}
\caption{(Color online) 
Spatially and temporally averaged nearest-neighbor density-density correlation $\langle \langle n_i n_j \rangle \rangle$ for vertical $ \mbox{\boldmath $r$}_{ij}$ as a function of polarization of photoexcitation $\theta$ for $U$=0.8, $V_1$=0.3, (a) $V_2$ = 0.25, (b) $V_2$ = 0.35, and (c) $V_2$ = 0.4, and different field amplitudes $ eaF/(\hbar\omega) $. 
(d) Spatially averaged nearest-neighbor density-density correlation of ground state $\langle n_i n_j \rangle$ for vertical $ \mbox{\boldmath $r$}_{ij}$ as a function of $V_2$ for $U$=0.8 and $V_1$=0.3. 
\label{fig:avnn5}}
\end{figure}

Next we show the spatially and temporally averaged nearest-neighbor density-density correlation $\langle \langle n_i n_j \rangle \rangle$. For $V_2=0.35$, which is slightly larger than $V_1=0.3$, we know that the anisotropy in the effective intersite repulsive interactions is enhanced by photoexcitation,\cite{yonemitsu_jpsj17a} which contradicts Eq.~(\ref{eq:delta_V}). Then, we use $V_2=0.25$ and show $\langle \langle n_i n_j \rangle \rangle$ for non-vertical bonds in Fig.~\ref{fig:avnn1234}(a) as a function of $ \theta $. 
It increases as if the intersite repulsion $V_1$ were transiently decreased relative to the bandwidth after photoexcitation. Then, we vary $ V_1 $ and calculate the spatially averaged nearest-neighbor density-density correlation of the ground state $\langle n_i n_j \rangle$ for non-vertical bonds, as shown in Fig.~\ref{fig:avnn1234}(b). For $ \theta = 0 $ and $ eaF/(\hbar\omega) = 0.4 $ in Fig.~\ref{fig:avnn1234}(a), $\langle \langle n_i n_j \rangle \rangle$ is about 0.17. To reproduce this value in the ground state, we need to decrease $ V_1 $ by 8\%, as shown in Fig.~\ref{fig:avnn1234}(b). This is not explained by the lowest-order effect shown in Fig.~\ref{fig:flq_hfe}(a). Relative interaction strengths are effectively and differently modulated as shown in Figs.~\ref{fig:avdo} and \ref{fig:avnn1234} and in Fig.~\ref{fig:avnn5} later. We need at least a second-order effect because in the high-frequency expansion different interactions are modulated differently from the second order. Figure~\ref{fig:flq_hfe}(b) shows the right-hand side of Eq.~(\ref{eq:2nd_order}), which implies that the modulation is about 1.5\%. Because the actual modulation depends on the other interaction parameters in Eq.~(\ref{eq:delta_V}) and there are effective interactions that are not in the form of density-density interactions, the discrepancy appears to be not so large. 

Finally we show $\langle \langle n_i n_j \rangle \rangle$ for vertical bonds as a function of $ \theta $ for $V_2=0.25$ in Fig.~\ref{fig:avnn5}(a), for $V_2=0.35$ in Fig.~\ref{fig:avnn5}(b), and for $V_2=0.4$ in Fig.~\ref{fig:avnn5}(c). 
It decreases for $V_2=0.25 < V_1$ as if the intersite repulsion $V_2$ were increased, increases for $V_2=0.4 > V_1$ as if $V_2$ were decreased, and therefore behaves as if the difference between $V_1$ and $V_2$ were suppressed. For $V_2=0.35$, as previously reported,\cite{yonemitsu_jpsj17a} the anisotropy in the effective $ V_{ij} $ is enhanced by photoexcitation, so that it is not described by the present effective Hamiltonian and its $ \theta $ dependence is different from that in Figs.~\ref{fig:flq_hfe}(a) and \ref{fig:flq_hfe}(b). We vary $ V_2 $ and calculate the ground state $\langle n_i n_j \rangle$ for vertical bonds, as shown in Fig.~\ref{fig:avnn5}(d). For $ \theta = 0 $ and $ eaF/(\hbar\omega) = 0.4 $, $\langle \langle n_i n_j \rangle \rangle$ is about 0.222 in Fig.~\ref{fig:avnn5}(a), which corresponds to an increase in $V_2$ in the ground state by 8\% (from $V_2=0.25$ to $V_2=0.27$) in Fig.~\ref{fig:avnn5}(d). At the same photoexcitation, $\langle \langle n_i n_j \rangle \rangle$ is about 0.13 in Fig.~\ref{fig:avnn5}(c), which corresponds to a decrease in $V_2$ in the ground state by 5\% (from $V_2=0.4$ to $V_2=0.38$) in Fig.~\ref{fig:avnn5}(d). These values (8\% and 5\%) are close to that for the required modulation of $V_1$ (8\%) above. The discrepancy from Fig.~\ref{fig:flq_hfe}(b) is not very large. More importantly, the anisotropy in the effective $ V_{ij} $ is reduced by photoexcitation in Figs.~\ref{fig:avnn5}(a) and \ref{fig:avnn5}(c), which is consistent with Eq.~(\ref{eq:delta_V}), and its $ \theta $ dependence is similar to that in Fig.~\ref{fig:flq_hfe}(b). 

\section{Conclusions and Discussion}

We have compared a continuous-wave-induced phenomenon with a pulse-induced one. For continuous waves, we employ a high-frequency expansion in the framework of quantum Floquet theory to obtain effective transfer integrals and interactions for extended Hubbard models, where the strengths and ranges of density-density interactions are arbitrary. The effective Hamiltonian is in principle valid only for the stroboscopic time evolution in steps of the period. Polarization dependences of the effective model parameters are focused on because polarization dependences are experimentally accessible through reflectivity spectra for instance. For monocycle pulses, we use the quarter-filled extended Hubbard model with nearest-neighbor interactions on a triangular lattice with linear trimers, which was previously used with particular parameters to explain the photoinduced enhancement of anisotropic charge correlations in $\alpha$-(BEDT-TTF)$_2$I$_3$ in the metallic phase. Modulations of correlation functions are investigated and their compatibility with the effective parameters is studied. 

The effective interactions for continuous waves are produced or modulated through double commutators by the square of a transfer integral multiplied by a non-zeroth-order Bessel function divided by the frequency of the field, as already known.\cite{rahav_pra03,mananga_jcp11,goldman_prx14,eckardt_njp15,itin_prl15,bukov_ap15,mikami_prb16} This knowledge is applied to density-density interactions with arbitrary strengths and ranges. New interaction terms that are absent in the original model contain electron transfers from or to a site that is included in the original interaction or linked by an electron transfer to another site included in the original interaction, as shown in the Appendix. They are either in the form of interactions between site-diagonal and site-off-diagonal densities or in the form of interactions between site-off-diagonal densities. They will generally tend to homogenize the electron distribution if the charge is initially disproportionated. Interactions between site-diagonal densities $ U_{i} $ and $ V_{ij} $ are modified in such a manner that large interaction parameters become small while small interaction parameters become large, thus averaging themselves out. Because the rates of modulations are governed by the square of $ t_{ij} J_m(ij) / \omega $, the polarization along the largest electron transfer $ \mid t_{ij} \mid $ is generally the most efficient. 

Numerical calculations are performed for time evolutions of correlation functions after monocycle pulse excitation on the basis of the exact diagonalization method. The time-averaged double occupancy behaves as if the on-site repulsion were increased relative to the bandwidth and its polarization dependence is comparable with the lowest-order effect in the high-frequency expansion. Except for the case where intersite repulsions compete, the time-averaged nearest-neighbor density-density correlations behave as if the anisotropy in intersite repulsions were averaged out. Their polarization dependences are comparable to the effect in the order of $ \omega^{-2} $ in the high-frequency expansion, although quantitative comparisons are difficult owing to many new effective interaction terms that contain electron transfers. 

The situations so far known and clarified here are summarized as follows. i) Immediately after monocycle pulse excitation, early-stage changes in the electronic state are similar to those after a sudden application of a continuous wave. ii) When we compare between time averages of correlation functions sufficiently after monocycle pulse excitation and those after a sudden application of a continuous wave with the same amplitude, the deviations from the corresponding quantities in the ground state are generally larger (but not much larger for high-frequency driving) for continuous waves. iii) A sudden application of a continuous wave (so-called ``ac quench'') and the corresponding interaction quench show very similar time evolutions of correlation functions if their rapid time variations in the former are averaged over the timescale of $ T $.\cite{tsuji_prl11} iv) For interaction quench from noninteracting to interacting many-body systems, particular nonequilibrium expectation values are twice as large as their corresponding analogues in equilibrium.\cite{moeckel_prl08,moeckel_anphys09} Such overshoot phenomena are expected for interaction quench between finite strengths and also for ac quench. 

From all these comparisons, we find that the effective Hamiltonian is useful in roughly predicting tendencies in correlation functions after monocycle pulse excitation. However, the effective Hamiltonian is independent of the filling or the system size, so that it is not directly be applicable to phenomena particular to a special filling or highly nonlinear phenomena such as the photoinduced enhancement of anisotropic charge correlations in $\alpha$-(BEDT-TTF)$_2$I$_3$ in the metallic phase. 

\begin{acknowledgments}
The author is grateful to S. Iwai and Y. Tanaka for various discussions. 
This work was supported by Grants-in-Aid for Scientific Research (C) (Grant No. 16K05459) and Scientific Research (A) (Grant No. 15H02100) from the Ministry of Education, Culture, Sports, Science and Technology of Japan. 
\end{acknowledgments}

\appendix
\section{Double Commutators in Effective Hamiltonian}

The first double commutator appearing in Eq.~(\ref{eq:interaction_3UV}) is calculated as 
\begin{equation}
[ H_{-m},[ H_{U}, H_{m}]] = \sum_{l_1,l_2,l_3,\sigma} 
(A_{l_1,l_2,l_3,\sigma}+B_{l_1,l_2,l_3,\sigma}+C_{l_1,l_2,l_3,\sigma})
\;, \label{eq:H_mUm}
\end{equation}
where 
\begin{eqnarray}
A_{1,2,3,\sigma} & & = 
[t_{12}J_{-m}(12)t_{23}J_{m}(23)c^\dagger_{1,\sigma}c_{3,\sigma} \nonumber \\
& & + t_{32}J_{m}(32)t_{21}J_{-m}(21)c^\dagger_{3,\sigma}c_{1,\sigma}]
U_2 n_{2,-\sigma}
\;,
\end{eqnarray}
\begin{eqnarray}
B_{1,2,3,\sigma} & & = - 
[t_{12}J_{-m}(12)t_{23}J_{m}(23)c^\dagger_{1,\sigma}c_{3,\sigma} \nonumber \\
& & + t_{32}J_{m}(32)t_{21}J_{-m}(21)c^\dagger_{3,\sigma}c_{1,\sigma}]
U_3 n_{3,-\sigma}
\;,
\end{eqnarray}
and 
\begin{eqnarray}
& & C_{1,2,3,\sigma} = \nonumber \\
& &        \left[ t_{12}J_{-m}(12) c^\dagger_{1,\sigma} c_{2,\sigma} 
- t_{21}J_{-m}(21) c^\dagger_{2,\sigma} c_{1,\sigma}  \right] U_2 \nonumber \\
& & \times \left[ t_{23}J_{m}(23) c^\dagger_{2,-\sigma} c_{3,-\sigma} 
- t_{32}J_{m}(32) c^\dagger_{3,-\sigma} c_{2,-\sigma} \right] 
\;.
\end{eqnarray}
If we write the $l_1 = l_3 $ and $ l_1 \neq l_3 $ terms separately and substitute them into Eq.~(\ref{eq:interaction_3UV}), we reproduce Ref.~\citen{itin_prl15} for the case of the one-dimensional Hubbard model with homogeneous interaction strengths. 

The second double commutator appearing in Eq.~(\ref{eq:interaction_3UV}) is calculated as 
\begin{equation}
[ H_{-m},[ H_{V}, H_{m}]] = \sum_{l_0,l_1,l_2,l_3,\sigma,\tau} 
(A_{l_0,l_1,l_2,l_3,\sigma,\tau}
+B_{l_0,l_1,l_2,l_3,\sigma,\tau}
+C_{l_0,l_1,l_2,l_3,\sigma,\tau})
\;, \label{eq:H_mVm}
\end{equation}
where 
\begin{eqnarray}
A_{0,1,2,3,\sigma,\tau} & & = 
[t_{12}J_{-m}(12)t_{23}J_{m}(23)c^\dagger_{1,\sigma}c_{3,\sigma} \nonumber \\
& & + t_{32}J_{m}(32)t_{21}J_{-m}(21)c^\dagger_{3,\sigma}c_{1,\sigma}]
V_{20} n_{0,\tau}
\;,
\end{eqnarray}
\begin{eqnarray}
B_{0,1,2,3,\sigma,\tau} & & = - 
[t_{12}J_{-m}(12)t_{23}J_{m}(23)c^\dagger_{1,\sigma}c_{3,\sigma} \nonumber \\
& & + t_{32}J_{m}(32)t_{21}J_{-m}(21)c^\dagger_{3,\sigma}c_{1,\sigma}]
V_{30} n_{0,\tau}
\;,
\end{eqnarray}
and 
\begin{eqnarray}
& & C_{0,1,2,3,\sigma,\tau} = \nonumber \\
& & \left[ t_{01}J_{-m}(01) c^\dagger_{0,\sigma} c_{1,\sigma} 
- t_{10}J_{-m}(10) c^\dagger_{1,\sigma} c_{0,\sigma}  \right] V_{12} \nonumber \\
& & \times \left[ t_{23}J_{m}(23) c^\dagger_{2,\tau} c_{3,\tau} 
- t_{32}J_{m}(32) c^\dagger_{3,\tau} c_{2,\tau} \right] 
\;.
\end{eqnarray}
If we separately write the terms where some of the $l_0$, $l_1$, $l_2$, and $l_3$ are equal, we have, in addition to the above (different numbers denote different sites now), 
\begin{eqnarray}
& & \left. A_{0,1,2,3,\sigma,\tau} \right|_{l_1=l_3}
+ \left. B_{0,1,2,3,\sigma,\tau} \right|_{l_1=l_3} =  \nonumber \\
& & [t_{32}J_{-m}(32)t_{23}J_{m}(23) 
+ t_{32}J_{m}(32)t_{23}J_{-m}(23)] \nonumber \\
& & \times n_{3,\sigma} (V_{20}-V_{30}) n_{0,\tau}
\;, \label{eq:density_density_1}
\end{eqnarray}
\begin{eqnarray}
& & \left. A_{0,1,2,3,\sigma,\tau} \right|_{l_1=l_0}
+ \left. B_{0,1,2,3,\sigma,\tau} \right|_{l_1=l_0} =  \nonumber \\
& & [t_{02}J_{-m}(02)t_{23}J_{m}(23)c^\dagger_{0,\sigma}c_{3,\sigma} \nonumber \\
& & + t_{32}J_{m}(32)t_{20}J_{-m}(20)c^\dagger_{3,\sigma}c_{0,\sigma}] 
(V_{20}-V_{30}) n_{0,\tau}
\;,
\end{eqnarray}
\begin{eqnarray}
& & \left. A_{0,1,2,3,\sigma,\tau} \right|_{l_3=l_0} =  \nonumber \\
& & [t_{12}J_{-m}(12)t_{20}J_{m}(20)c^\dagger_{1,\sigma}c_{0,\sigma} \nonumber \\
& & + t_{02}J_{m}(02)t_{21}J_{-m}(21)c^\dagger_{0,\sigma}c_{1,\sigma} ] 
V_{20} n_{0,\tau}
\;,
\end{eqnarray}
\begin{eqnarray}
& & \left. B_{0,1,2,3,\sigma,\tau} \right|_{l_2=l_0} = \nonumber \\
& & -[t_{10}J_{-m}(10)t_{03}J_{m}(03)c^\dagger_{1,\sigma}c_{3,\sigma} \nonumber \\
& & + t_{30}J_{m}(30)t_{01}J_{-m}(01)c^\dagger_{3,\sigma}c_{1,\sigma}]
V_{30} n_{0,\tau}
\;,
\end{eqnarray}
\begin{eqnarray}
& & \left. C_{0,1,2,3,\sigma,\tau} \right|_{l_0=l_3} = \nonumber \\
& & \left[ t_{31}J_{-m}(31) c^\dagger_{3,\sigma} c_{1,\sigma} 
- t_{13}J_{-m}(13) c^\dagger_{1,\sigma} c_{3,\sigma}  \right] V_{12} \nonumber \\
& & \times \left[ t_{23}J_{m}(23) c^\dagger_{2,\tau} c_{3,\tau} 
- t_{32}J_{m}(32) c^\dagger_{3,\tau} c_{2,\tau} \right]
\;,
\end{eqnarray}
\begin{eqnarray}
& & \left. C_{0,1,2,3,\sigma,\tau} \right|_{l_0=l_2} = \nonumber \\
& & \left[ t_{21}J_{-m}(21) c^\dagger_{2,\sigma} c_{1,\sigma} 
- t_{12}J_{-m}(12) c^\dagger_{1,\sigma} c_{2,\sigma}  \right] V_{12} \nonumber \\
& & \times \left[ t_{23}J_{m}(23) c^\dagger_{2,\tau} c_{3,\tau} 
- t_{32}J_{m}(32) c^\dagger_{3,\tau} c_{2,\tau} \right]
\;,
\end{eqnarray}
\begin{eqnarray}
& & \left. C_{0,1,2,3,\sigma,\tau} \right|_{l_3=l_1} = \nonumber \\
& & \left[ t_{01}J_{-m}(01) c^\dagger_{0,\sigma} c_{1,\sigma} 
- t_{10}J_{-m}(10) c^\dagger_{1,\sigma} c_{0,\sigma}  \right] V_{12} \nonumber \\
& & \times \left[ t_{21}J_{m}(21) c^\dagger_{2,\tau} c_{1,\tau} 
- t_{12}J_{m}(12) c^\dagger_{1,\tau} c_{2,\tau} \right]
\;,
\end{eqnarray}
\begin{eqnarray}
& & \left. A_{0,1,2,3,\sigma,\tau} \right|_{l_0=l_1=l_3=l_i; l_2=l_j}
+ \left. B_{0,1,2,3,\sigma,\tau} \right|_{l_1=l_3=l_i; l_0=l_2=l_j} =  \nonumber \\
& & [t_{ij}J_{-m}(ij)t_{ji}J_{m}(ji) 
+ t_{ij}J_{m}(ij)t_{ji}J_{-m}(ji)] \nonumber \\
& & \times \frac{1}{2} (n_{i,\sigma}-n_{j,\sigma}) 
V_{ij} (n_{i,\tau}-n_{j,\tau})
\;, \label{eq:density_density_2}
\end{eqnarray}
and 
\begin{eqnarray}
& & \left. C_{0,1,2,3,\sigma,\tau} \right|_{l_0=l_2, l_3=l_1} = \nonumber \\
& & \left[ t_{21}J_{-m}(21) c^\dagger_{2,\sigma} c_{1,\sigma} 
- t_{12}J_{-m}(12) c^\dagger_{1,\sigma} c_{2,\sigma}  \right] V_{12} \nonumber \\
& & \times \left[ t_{21}J_{m}(21) c^\dagger_{2,\tau} c_{1,\tau} 
- t_{12}J_{m}(12) c^\dagger_{1,\tau} c_{2,\tau} \right]
\;.
\end{eqnarray}
Because $ [ H_{-m},[ H_{U}, H_{m}]] $ is a special case of $ [ H_{-m},[ H_{V}, H_{m}]] $ (i.e., $ U_{i} = V_{ii} $), we omit the corresponding terms for $ [ H_{-m},[ H_{U}, H_{m}]] $. 

\bibliography{68007}

\end{document}